\documentclass[twocolumn,3p]{elsarticle}
\usepackage{graphicx}
\usepackage{pstricks}
\usepackage{dcolumn}
\usepackage{bm}
\usepackage{epsf}
\usepackage{epsfig}
\usepackage{amsmath,amsfonts,amsthm, amssymb,latexsym}
\usepackage{enumerate}

\newcommand{\beq}{\begin{equation}}
\newcommand{\eeq}{\end{equation}}
\newcommand{\bcn}{\begin{center}}
\newcommand{\ecn}{\end{center}}

\newcommand{\Omg}{\Omega}

\newcommand{\msun}{$\, M_\odot$}

\newcommand{\lsim}{\lower0.5ex\hbox{$\; \buildrel < \over \sim \;$}}

\journal{Physics Letters B}
\begin{document}
\begin{frontmatter}
\title{Impact of Rotation-Driven Particle Repopulation on the Thermal Evolution
of Pulsars}

\date{\today}

\author[UFF,FIAS]{Rodrigo Negreiros} 
\ead{negreiros@fias.uni-frankfurt.de}
 
\author[FIAS]{Stefan Schramm}
\ead{schramm@th.physik.uni-frankfurt.de}

\author[SDSU]{Fridolin Weber}
\ead{fweber@mail.sdsu.edu}

 \address[UFF]{Instituto de Fisica,
Universidade Federal Fluminense,
Av. Gal. Milton Tavares de Souza s/n, Gragoata, Niteroi, 24210-346, Brazil
 }

\address[FIAS]{FIAS, Goethe University, Ruth Moufang Str. 1
         60438 Frankfurt, Germany
 }

\address[SDSU]{Department of Physics, San Diego State University, 5500
  Campanile Drive, San Diego, California 92182, USA}

\begin{abstract}
  Driven by the loss of energy, isolated rotating neutron stars
  (pulsars) are gradually slowing down to lower frequencies, which
  increases the tremendous compression of the matter inside of
  them. This increase in compression changes both the global
  properties of rotating neutron stars as well as their hadronic core
  compositions. Both effects may register themselves observationally
  in the thermal evolution of such stars, as demonstrated in this
  Letter. The rotation-driven particle process which we consider here
  is the direct Urca (DU) process, which is known to become operative
  in neutron stars if the number of protons in the stellar core
  exceeds a critical limit of around 11\% to 15\%. We find that
  neutron stars spinning down from moderately high rotation rates of a
  few hundred Hertz may be creating just the right conditions where
  the DU process becomes operative, leading to an observable effect
  (enhanced cooling) in the temperature evolution of such neutron
  stars. As it turns out, the rotation-driven DU process could explain
  the unusual temperature evolution observed for the neutron star in
  Cas~A, provided the mass of this neutron star lies in the range of
  1.5 to 1.9 \msun~ and its rotational frequency at birth was between
  40 (400 Hz) and 70\% (800 Hz) of the Kepler (mass shedding)
  frequency, respectively.
\end{abstract}

\end{frontmatter}



Numerical cooling simulations of neutron stars allow one to probe the
inner structure of these objects, and the properties of ultra-dense
matter
\cite{Schaab1996,Page2004,Page2006,Page2009,Blaschke2000,Grigorian2005,
  Blaschke2006,Page2011,Yakovlev2011}.  In the standard treatment,
neutron star cooling calculations are carried out for stellar core
compositions which are frozen-in, that is, compositions which do not
change with time, as it is the case for non-rotating neutron stars.
The situation may be very different for isolated rotating neutron
stars, which are spinning down due to magnetic braking. Such
stars can experience drastic density changes during spin-down, causing
the formation of novel states of matter in their cores. Examples of
which are the formation of a mixed phase of quarks and hadrons, of
pure quark matter, or a condensate made of bosons
\cite{Weber,Glendenning2000,page06:review,sedrakian07:a,alford08:a}.
Another intriguing possibility concerns the rotation-driven changes in
the number densities of neutrons, protons and leptons in the cores of
rotating neutron stars, which can have important observable
consequences for the thermal evolution of such objects, as shown in
this Letter.

Computing the thermal evolution of rotating neutron stars is
considerably more complicated than computing the cooling of
non-rotating neutron stars. The reasons are several. First, stellar
rotation requires solving Einstein's field equations for rotationally
deformed fluid distributions \cite{Weber,Glendenning2000}, which
renders the problem 2-dimensional. Second, the general relativistic
frame dragging effect imposes an additional self-consistency condition
on Einstein's field equations. Third, an extra self-consistency
condition is encountered when calculating the general relativistic
Kepler (mass shedding) frequency of a rotating neutron star.  Fourth,
the thermal transport equations need to be solved for general
relativistic, non-spherical fluids that may experience anisotropic
heat transport. Our study accounts for the first three effects. The
fourth effect, the anisotropic transport of heat inside neutron stars,
is reserved for a separate future study \cite{Negreiros2011}.

In this Letter, we consider the cooling behavior of isolated rotating
neutron stars. The number of baryons (i.e., the so-called baryon mass)
of such stars remains unchanged during spin-down.  
As known from earlier studies \cite{Weber,Weber2005,Negreiros2010},
the spin-down driven gravitational compression can cause substantial
changes in the core compositions of such neutron stars.  A striking
example of this is the direct Urca (DU) process, which describes the
\begin{figure}[tb]
  \includegraphics[scale=0.30]{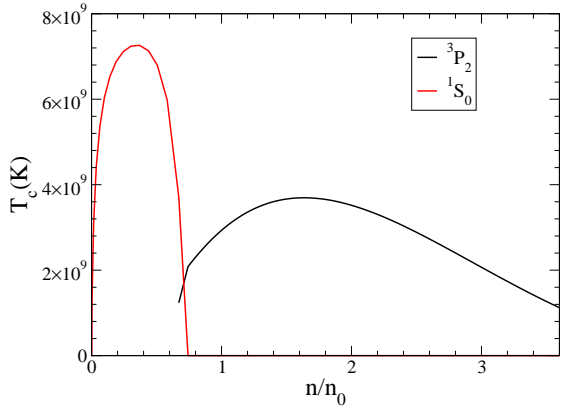}
  \caption{(Color online) Critical temperature of superfluid neutrons
    in $^1S_0$ and $^3P_2$ states as a function of baryon number
    density. ($n_0$ denotes the density of ordinary nuclear matter.)}
\label{SF} 
\end{figure}
direct (no bystander particle necessary) transformation of neutrons to
protons, electrons and anti-electron neutrinos in the cores of neutron
stars, according to
\begin{eqnarray}
  n \rightarrow  p + e^- + \bar\nu_e \, ,
\label{eq:DU}
\end{eqnarray}
Reaction (\ref{eq:DU}) can only occur if the proton fraction exceeds
between 11\% to 15\% \cite{Lattimer1991}.  The neutrino luminosity
associated with this reaction dominates over those of other neutrino
emitting processes in the core \cite{Page2006,Lattimer1991}. One may
thus speculate that once (\ref{eq:DU}) becomes operative in rotating
neutron stars, it could speed up the cooling of such stars
substantially, leading to an observational signature in the thermal
evolution of such objects.  The pairing of neutrons in the $^1S_0$
singlet and $^3P_2$ triplet states, which reduces the neutrino
emission from neutron stars, would not change this behavior, as will
be discussed later in this Letter.  Following the approach of
\cite{Schaab94,Levenfish94}, the critical temperatures assumed for
these states are shown in Fig.\ \ref{SF}.

\begin{figure*}[htb]
\begin{center}
\begin{tabular}{cc}
 \includegraphics[width=0.450\textwidth]{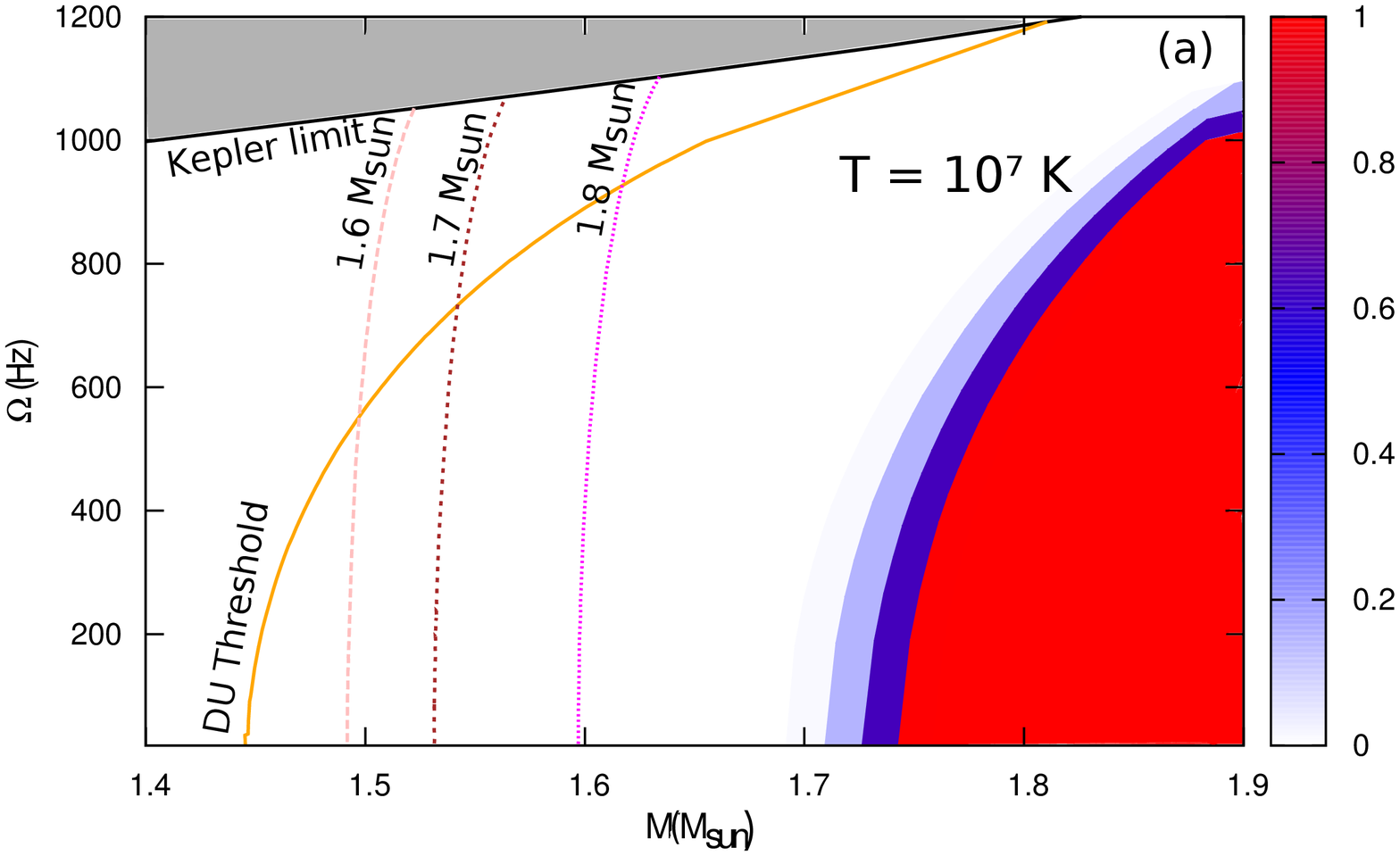} ~~&
 \includegraphics[width=0.450\textwidth]{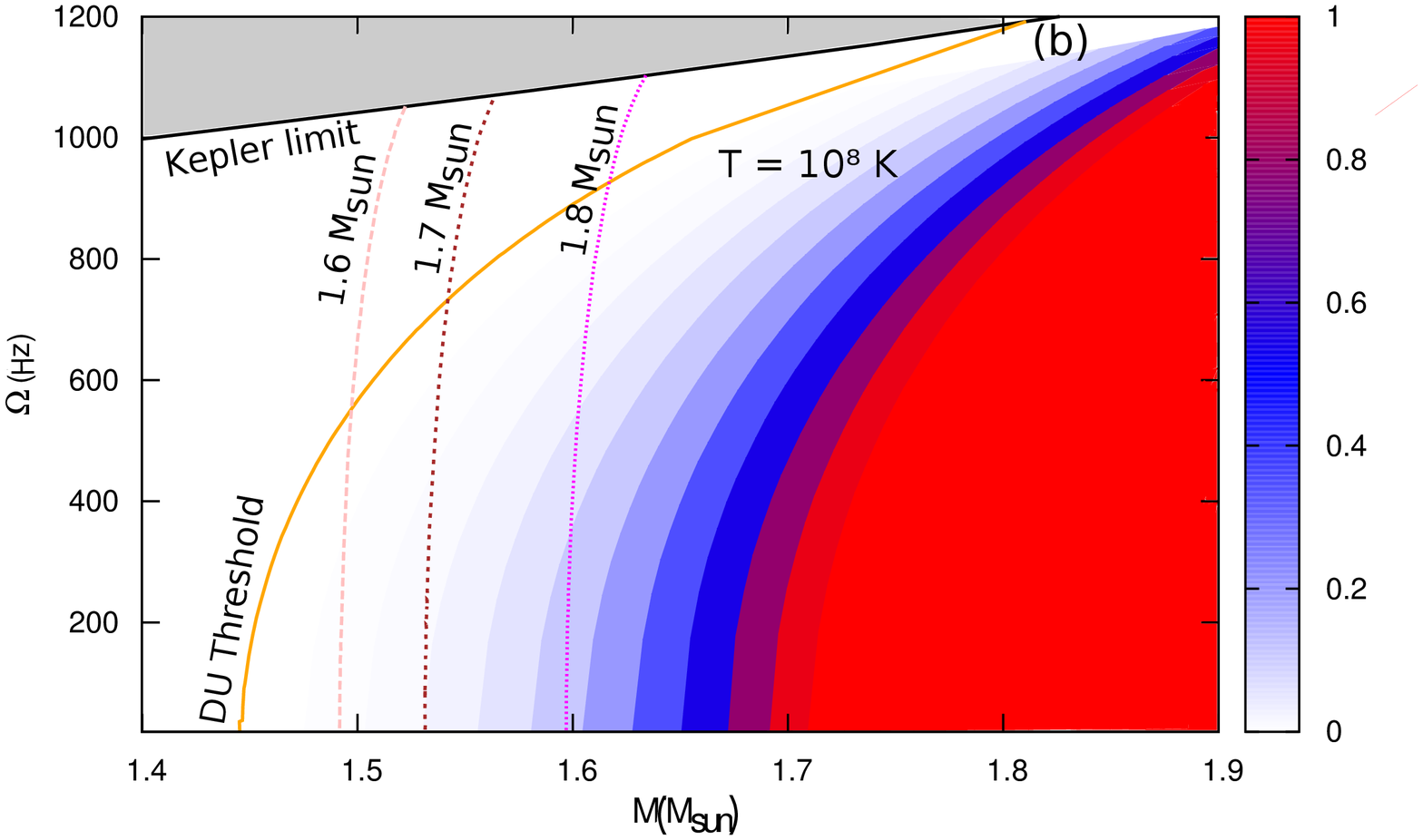} \\
& \\
 & \\
 & \\
\includegraphics[width=0.450\textwidth]{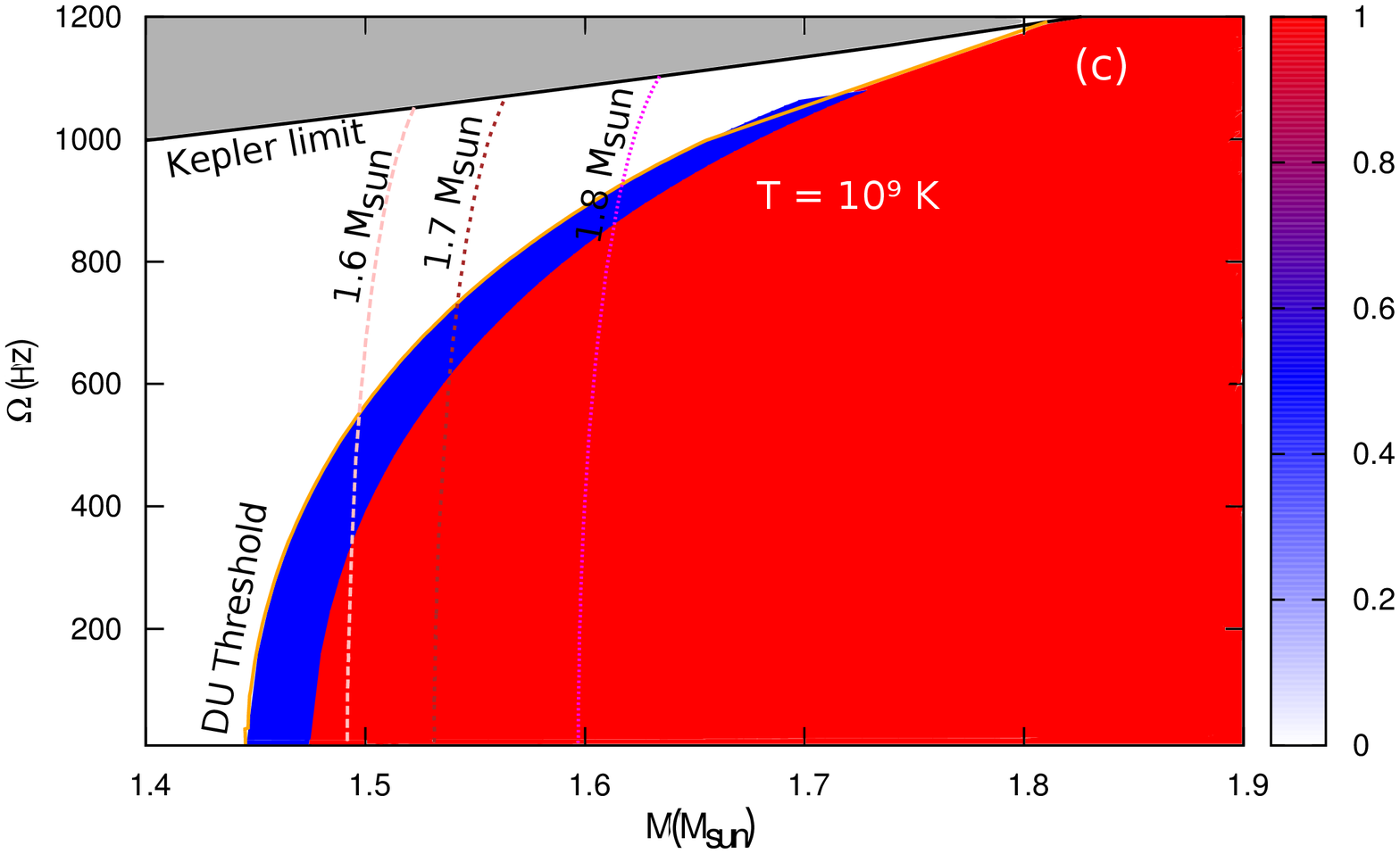} ~~&
\includegraphics[width=0.450\textwidth]{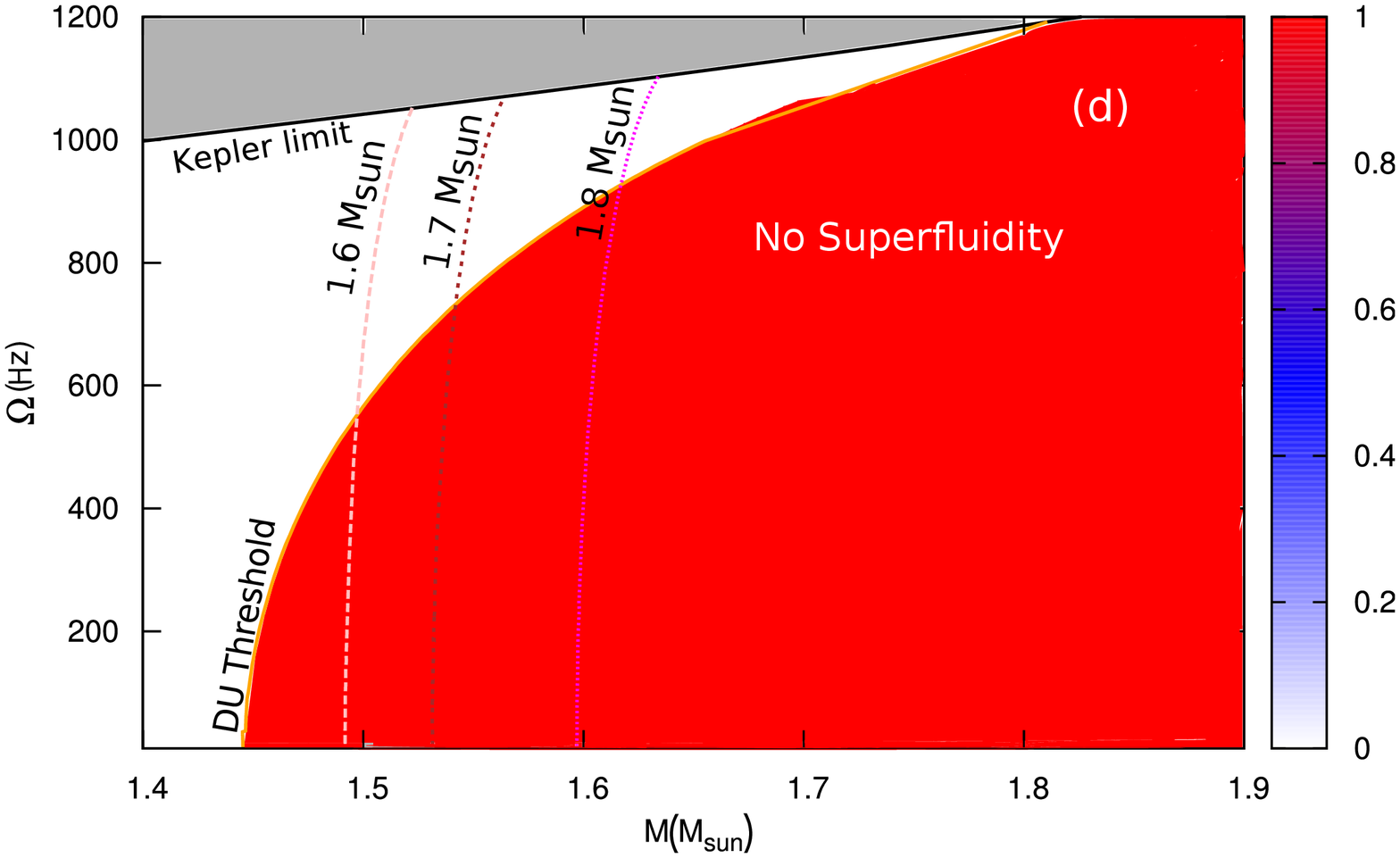}
\end{tabular}
\caption{(Color online) Intensity of the direct Urca (DU) process in
  rotating (frequency $\Omega$) neutron stars, for different stellar
  temperatures, $T$. $M$ denotes the gravitational mass.  The
  intensity ranges from 0 (plain white areas) to 100\% (red). No stars
  are allowed above the curve labeled Kepler limit because of stellar
  mass shedding.  The vertical curves show evolutionary tracks of
  isolated rotating neutron stars, whose baryon mass ($M_0$) during
  spin-down remains constant. One sees that for several stars evolving
  along these tracks the DU process (i.e., reaction (\ref{eq:DU}))
  becomes possible at lower rotation rates. The color coding indicates
  the intensity of the DU process.}
  \label{Omgxfreq}
\end{center}
\end{figure*}

Considering the conservation of momentum and the superfluidity
suppression, we show in Fig.\ \ref{Omgxfreq} the intensity of the DU
process (defined here as the reduction factor introduced by pairing,
i.e.\ ``1'' means no reduction, ``0'' total suppression) in the core
of rotating neutron stars for different stellar masses, $M$, and
rotational frequencies, $\Omega$.  One sees that, depending on
frequency (and thus central density) and temperature (which affects
the level at which the DU process is suppressed), the DU process
operates at very different intensities.  Hence, as a rotating neutron
star evolves from a given ``initial'' state (given frequency, mass,
and temperature) in time, the intensity of the DU process at its core
can vary substantially. Also shown in Fig.\ \ref{Omgxfreq} are several
evolutionary tracks of neutron stars which spin down because of the
emission of magnetic dipole radiation. The intensity of the DU process
of stars along these tracks can vary significantly too, depending on
the values of the star's initial parameters, which ought to affect
their thermal evolution.

In passing we note that the rotationally driven compositional changes
in the cores of rotating neutron stars may also occur, in reverse
however, in the cores of neutron stars in binaries, which are spun up
by mass accretion. The treatment of such neutron stars is more
complicated, however, since both the frequency as well as the star's
mass are changing.  In that case the evolutionary tracks shown in
Fig.~\ref{Omgxfreq} will follow a different path, but the overall
modification of the intensity of the DU process may be expected to
hold.

The results shown in Fig.\ \ref{Omgxfreq} are generic for any EoS
which predicts a sufficiently large number of protons in the cores of
neutron stars so that reaction (\ref{eq:DU}) becomes possible.
The EoS used in this Letter is a parametrized version of the
Akmal-Pandharipande-Ravenhall (APR) EoS
\cite{Akmal1998,Heiselberg2000} (and references therein). For this
parametrization, the binding energy consists of a compression term and
a symmetry energy term,
\begin{eqnarray}
  BE = E_0  \ u  \  \frac{u - 2 -\delta }{1 + \delta u} + S_0 \ u^\gamma \ 
  (1 - 2 x_p)^2  \, ,
\label{eq:eos}
\end{eqnarray}
where $E_0$ is the saturation energy, $S_0$ the symmetry energy at
saturation, $u \equiv n/n_0$ with $n$ the baryon number density and
$n_0$ the baryon number density at saturation, and $x_p$ the proton
fraction. The parameters $\delta$ and $\gamma$ can be used to control
the nuclear incompressibility and the symmetry energy,
respectively. The maximum mass of a neutron star is thus sensitive to
changes in $\delta$, and the threshold density for the onset of the DU
process is sensitive to changes in $\gamma$. Details of the
parametrization of the APR EoS have been discussed in
\cite{Heiselberg2000}.  The cooling of neutron stars described by this
EoS were first explored in \cite{Gusakov2005}. Very recently, this EoS
was used to put constraints on the properties (superfluid gaps) of
ultra-dense matter using the temperature data observed
for the neutron star in Cas~A
\cite{Page2011,Yakovlev2011,Shternin2011}.

Next, we explore the impact of a varying proton fraction, driven by
stellar rotation, on the thermal evolution of isolated rotating
neutron stars.  In this study we consider the following cooling
processes: the direct Urca, the modified Urca, and Bremsstrahlung
processes in the stellar core. For the crust, we consider
Bremsstrahlung, $e^+e^-$ pair annihilation, and plasmon decay
processes. A comprehensive overview of the neutrino emission processes
in neutron stars can be found in
\cite{Weber,page06:review,Yakovlev2001a}. The boundary conditions are
determined by the luminosity at the stellar center, $l(0) = 0$, and at
the surface $l(R) = L_S$ \cite{Gudmundsson1983,Potekhin1997}.  To
establish a connection between the rotational frequency of neutron
stars and their respective cooling stages, we model the spin-down rate
according to
\begin{eqnarray}
  \dot{\Omg} = -\ K \ \Omg^n \, ,
  \label{eq:n}
\end{eqnarray} 
where $\Omg$ denotes the neutron star spin frequency, and $K$ and $n$ are
constants \cite{Glendenning2000}.  The constant $K$ can be parametrized
as  (for $n=3$)
$K = 1.55\times 10^{-17}\times R_{10}^4\, B_{12}^2\, / (M/M_\odot)$s,
where $B_{12}$ is the surface magnetic field in units of $10^{12}$ G,
$R_{10}$ the radius in units of 10 km, and $\alpha$ the inclination
angle of $B$ (note that the moment of inertia was approximated by
$2MR^2/5$). Using the frequencies and time derivatives of observed
pulsars available at the ATNF pulsar data base \cite{Manchester2005a},
we find that $10^{-21} \leq K \leq 10^{-11}$ (in the same units as
above, i.e., s) for a braking index $n = 3$. The latter value
corresponds to neutron stars which spin-down due to the emission of
magnetic dipole radiation, as considered in this Letter. By
integrating Eq.~(\ref{eq:n}) one  obtains the frequency as a
function of time.  With the aid of Eq.\ (\ref{eq:n}), we can now trace
the thermal evolution of isolated, spinning-down neutron stars. 

\begin{figure}[htb]
  \includegraphics[trim = 0mm 1mm 0mm 0mm,
clip,scale=0.30,angle=0]{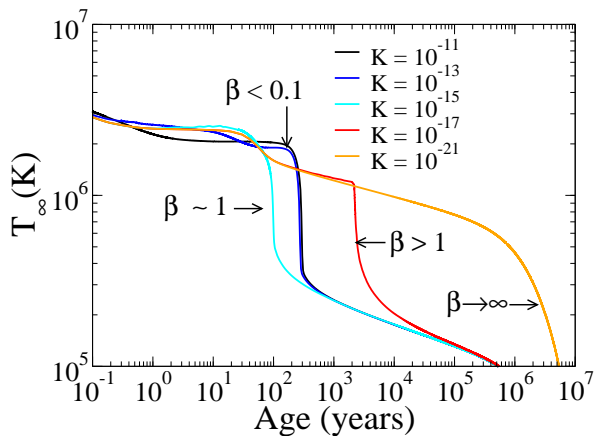}
\caption{(Color online) Neutron star temperature at infinity as a
  function of stellar age, for a sample neutron star with baryon mass
  of 1.5 \msun. (The gravitational mass at zero rotation is 1.4
  \msun.)  Depending on the value of $K$, four distinctly different
  cooling epochs emerge, which are discussed in the text.}
\label{T_x_age} 
\end{figure}

\begin{figure}[htb]
  \includegraphics[trim = 0mm 1mm 0mm
  0mm,clip,scale=0.30,angle=0]
{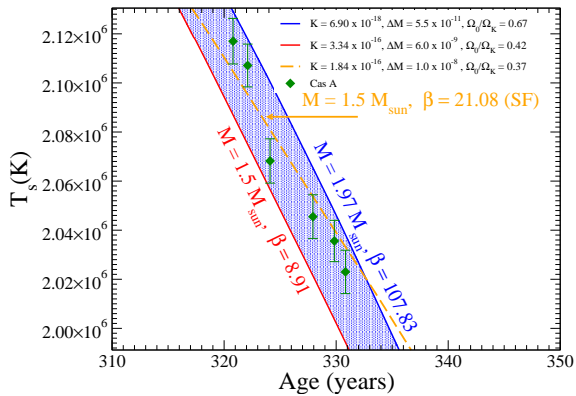}
\caption{(Color online) Cooling simulations reproducing the
  temperatures observed for the neutron star in Cas~A over a ten-year
  period.  $K$ is defined in the text, $\Delta M$ denotes the mass (in
  unit of the mass of the neutron star) of the accreted envelope, and
  $\Omega_0 / \Omega_K$ denotes the star's birth frequency relative to
  the Kepler frequency. Superfluidity and Pair-Breaking Formation of
  nucleons is taken into account in the curve labeled SF.}
\label{fig:CasA} 
\end{figure}

The results are shown in Figs.~\ref{T_x_age} and \ref{fig:CasA}.  It
is evident from Fig.~\ref{T_x_age} that the temperature drops
dramatically when the DU process (\ref{eq:DU}) becomes operative
during spin-down.  The characteristic time scale of the thermal
coupling between the core and the crust of non-rotating stars is
$\tau_C =(\Delta R/1\mbox{km})^2
(1-5.04\times10^{-2}(M/M_\odot)/R_{10})^{-3/2} t_1$, where $\Delta R$
is the crust thickness, and $t_1$ is a constant with dimension of
time, that depends on the microscopic properties of the star, and
whose value is $\sim 30$ years \cite{Gnedin2001}. For a traditional
neutron star $\tau_C = 50 - 150$ years \cite{Gnedin2001}, and in this
work we set $\tau_C = 100$~years . In the case of spinning-down
neutron stars considered here, we must also consider the relaxation
time of the spin-evolution, which we define as
the time it takes for the star to spin down to half of its original
frequency. This time scale is given by $\tau_S = 3/(2K\Omg_0^2)$,
where $\Omg_0$ denotes the birth frequency of the object.  The parameter
\begin{equation}
 \beta \equiv \frac{\tau_S}{\tau_C} = \frac{M/M_\odot (1 - 5.04\times10^{-2}
(M/M_{\odot})/R_{10})^{3/2} }{1.03\times 10^{-17}
R_{10}^4 B_{12}^2 \sin^2\alpha \Delta R_{\text{km}}^2  \Omega_0^2 t_1}
\end{equation}
can then be used to determine the impact of the spin-down compression
on the cooling of the star. Evidently this analysis is only valid for
objects that are born in the white regions of the figures shown in
(\ref{Omgxfreq}), and thus may cross-over into the DU region (red
shaded areas) during their evolution. For the EoS studied here, this
is limited to objects whose non-rotating gravitational masses are
$\gtrsim 1.5$ M$_\odot$.  The following cases emerge:
I.  If $\beta < 1$, the star spins down to low frequencies ($\Omg
\rightarrow 0$) before the core and crust are thermally coupled. This
causes the cooling of the star to be similar to the cooling of a
spherically symmetric star of same mass.  
II. If $\beta \sim 1$, the star is still rotating at relatively high
frequencies ($\sim \Omega_0/2$) when the core and crust become
thermally coupled.  The thicker crust of such a star allows the core
and crust to couple more quickly, effectively speeding up the cooling.
III. If $\beta > 1$, the star keeps rotating at high frequencies for a
long time, which delays the onset of the DU process (\ref{eq:DU}). We
note that, as was the case for $ \beta \sim 1$, the star also becomes
isothermal at an earlier time. However the late onset of the DU
process decouples the core and crust one more time, which yields to a
second thermal coupling, characterized by a sharp drop in temperature
at a later time.
IV. If $\beta \rightarrow \infty $, the spin-down relaxation time is much
greater
than the time scale of the core-crust thermal coupling. This means
that during most of the cooling period, the object will remain in the
high frequency domain, and the onset of the DU process may never be
achieved, leading to a slow stellar cooling.  The different scenarios
I through IV are graphically illustrated in Fig.~\ref{T_x_age}. The
curves underline the important role of $\beta$ for the classification
of the cooling behavior of spinning-down neutron star.  If superfluidity effects
were considered, one should expect that the temperature reductions 
for $\beta \sim 1$ and $\beta > 1$ are less pronounced. The actual
magnitude of the reduction will depend on the microscopic pairing
model used.

We conclude this Letter with a discussion of the temperature data
observed for the neutron star in Cas~A over a 10 year period
\cite{heinke2010}. The rapid cooling of this neutron star has been
explained recently through the presence of superfluidity in dense
matter \cite{Page2011, Yakovlev2011}. Our study indicates that the
temperature evolution of this neutron star could also be explained by
the late onset of the DU process, if one assumes that the neutron star
was created with an initial rotational frequency somewhere between 40
and 70\% of its mass shedding frequency (see Fig.\ \ref{fig:CasA}),
depending on the unknown mass of this neutron star.
Masses between 1.2 and 2.1 \msun \ (99\% confidence) have been
established for this neutron star \cite{Yakovlev2011}. The simulations
shown in Fig.\ \ref{fig:CasA} have thus been performed for two sample
masses, 1.5 and 1.9 \msun.  The mass of the accreted envelope, due to
the fallback of matter after the supernova explosion
\cite{Potekhin1997}, is assumed to be $\Delta M/M = 5.5\times10^{-11}$
and $6.0\times10^{-9}$, in accordance with \cite{ho2009}.  Our
calculations predict a birth frequency ($\Omega_0$) for the 1.5 \msun~
star of 42\% ($\Omega_ 0 = 400$~Hz) of the Kepler frequency if the
stars is made of non-superfluid matter. This figure drops down to 37\%
if neutron superfluidity in the $^1S_0$ and $^3P_2$ channels and
Pair-Breaking Formation \cite{Page2011} is included.  It is an open
issue whether proton superconductivity takes place in the core of a
neutron star. If it does, it may affect the thermal evolution of the
object \cite{Page2011}, depending on the magnitude and density
dependence of the superfluid gap and its critical temperature. In the
case that proton superconductivity extends to very high densities, and
has a high critical temperature, it might suppress the impact of the
DU process on the thermal evolution of a neutron star.

We note that on the basis of theoretical calculations of the mapping
between initial and final neutron star spins, likely spin-down
mechanisms, and observational constraints from pulsars and supernova
energetics, it has been argued \cite{Ott2006} that the initial
pre-collapse central iron core periods inside of massive stars, which
give birth to neutron stars and pulsars, are on average greater than
$\sim 50$ seconds.  The associated neutron stars would then be born
with rotation periods greater than around 10~ms \cite{Ott2006}. This
figure agrees with the birth periods established in
\cite{FaucherGiguere2006}.  If the central iron core periods should be
smaller by a factor of 5, the birth periods of neutron stars would
drop down to a millisecond. In our work, initial birth periods of
2.5~ms (400~Hz) and 1.25~ms (800~Hz) are required to fit the thermal
data of the neutron star in Cas A. Such rotational periods are on the
small end of probabilities, but given the many poorly understood
issues (estimated iron core spin rates, angular momentum profiles,
progenitor mass, general relativistic effects), which complicate the
study of stellar evolution with rotation, the birth periods considered
in our study are in the realms of possibility.

A further aspect concerns the energy loss during spin-down for the
assumed birth frequencies. For our model scenario, we find that
$\dot{E} = I\Omega \dot{\Omega} \sim 10^{39}$~erg/s. This figure would
imply that the neutron star in Cas A ought to emit a pulsar wind,
which, however, has not been observed.  If the pulsar wind should not
exist, the birth frequencies of the neutron star in Cas A may have
been somewhat smaller than the values assumed in this paper for
neutron stars made of non-superfluid matter. Interestingly, the
inclusion of superfluidity among neutrons and Pair-Breaking Formation
reduces the region in the star where the DU process takes place,
shifting the onset of the DU process toward lower rotational
frequencies (see curve labeled ``SF'' in Fig.\ \ref{fig:CasA}).

The results shown in Fig.~\ref{fig:CasA} were obtained by assuming
value of $K$ in the range of $\sim 10^{-16}$ and $\sim 10^{-18}$. The
value of $K$ is connected to the magnetic field through $K \propto
R_{10}^4 B_{12}^2/(M/M_\odot)$. It is known from observations that the
upper limit for the magnetic field of the neutron star in Cas A is
$2\times 10^{11}$~G, which would imply $K\sim 10^{-18}$ (assuming a
stellar mass of M =2.0 M$_\odot$ and a radius of $R = 15$~km), in
agreement with one of the parameter combinations considered in Fig.~
\ref{fig:CasA} (blue curve).

We also note that given the different nature of the processes employed
to explain the temperature drop (late onset of the DU process in our
case, and pair breaking/formation in references \cite{Page2011,
  Yakovlev2011}), continuous observation of the neutron star in Cas A
may decide which model is favorable.  A continuous analysis of the
slope of the temperature ($ s = d\log_{10}T_S/d\log_{10}t$) might
allow one to do that. While the slope found by us ($s = -1.5$) should
remain at this value while the surface temperature drops drastically,
in the model described in \cite{Page2011}, it should go to the
asymptotic value found by the authors ($s =-1/12$), before the surface
temperature changes drastically.

Our results show that rotation-driven repopulation processes can be of
very great importance for the thermal evolution of spinning-down
neutron stars. The reason for this is two-fold: first, the geometry of
the object is modified as they spin down, which changes the stellar
surface gravity and therefore the stellar surface temperature
\cite{Potekhin1997}. Second, the spin-down renders neutron stars
gradually more dense, which changes the stellar core compositions.  We
also discovered that the ratio of the spin-down relaxation time to the
core-crust coupling time, $\beta$, emerges as a most valuable
parameter that serves to classify the cooling behavior of rotating
pulsars. Furthermore, the methodology discussed in this Letter can be
generalized to more elaborated scenarios like accreting (X-ray)
neutron stars, and possibly objects undergoing more rapid braking
mechanisms like r-modes.  Evidently, in such cases, re-heating
processes are important, although they do not invalidate the
conclusions drawn by us in this Letter, i.e. that particle
repopulation plays an important role for the thermal evolution of
neutron stars.

In summary, the key point put forth in this letter is that the
particle composition in the core of an isolated rotating neutron star
changes due to the spin-down compression of the star caused by
magnetic braking. This compression is a robust physical phenomenon,
experienced by any isolated rotating neutron star.  This scenario is
applied to the neutron star in Cas A. We show that the remarkable
temperature drop of this neutron star can be explained by the onset of
the rotation-driven direct Urca process. We also show that continuous
observation of the thermal evolution of this neutron star over the
next ten to twenty years will allow us to determine whether our
explanation of the fast cooling of this neutron star is correct, or
whether other physical processes (as suggested in
\cite{Page2011,Yakovlev2011}) are responsible for the star's dramatic
temperature drop.

We acknowledge access to the computing facility of the Center of
Scientific Computing at the Goethe-University Frankfurt, where our
numerical calculations were performed.  F.\ W.\ is supported by the
National Science Foundation (USA) under Grant PHY-0854699.

\end{document}